\documentclass[prb,amsmath,amssymb,twocolumn,superscriptaddress]{revtex4-2}

\newcommand{\bra}[1] { \langle #1 | }
\newcommand{\ket}[1] { | #1 \rangle }
\newcommand{\drm}[1]{_{\mathrm{ #1}}}
\newcommand{\urm}[1]{^{\mathrm{ #1}}}

\usepackage{graphicx}                           
\usepackage{amsmath}                            
\usepackage{amssymb}                            
\usepackage{tabularx}                           
\usepackage{multirow}
\usepackage{hyperref}
\usepackage{array}                              
\usepackage{tikz}                               
\usepackage{mathrsfs}                           
\usepackage{mwe}                                
\usepackage{csquotes}

\begin{document}
\title{Plasmonic polarons induced by alkali-atom deposition in hafnium disulfide (1\textit{T}-HfS$_2$)}
\author{Christoph Emeis}
\affiliation{Institut für Theoretische Physik und Astrophysik, Kiel University, 24098 Kiel, Germany}
\author{Sanjoy Kr Mahatha}
\affiliation{UGC-DAE Consortium for Scientific Research, University Campus, Khandwa Road, Indore - 452001, India}
\affiliation{Ruprecht‑Haensel‑Labor, Deutsches Elektronen-Synchrotron DESY, 22607 Hamburg, Germany}
\author{Sebastian Rohlf}
\affiliation{Institut für Experimentelle und Angewandte Physik, Kiel University, 24098 Kiel, Germany}
\author{Kai Rossnagel}
\affiliation{Ruprecht‑Haensel‑Labor, Deutsches Elektronen-Synchrotron DESY, 22607 Hamburg, Germany}
\affiliation{Institut für Experimentelle und Angewandte Physik, Kiel University, 24098 Kiel, Germany}
\affiliation{Kiel Nano, Surface and Interface Science KiNSIS, 24118 Kiel, Germany}
\author{Fabio Caruso}
\affiliation{Institut für Theoretische Physik und Astrophysik, Kiel University, 24098 Kiel, Germany}
\affiliation{Kiel Nano, Surface and Interface Science KiNSIS, 24118 Kiel, Germany}
\date{\today}

\begin{abstract}
We combine ab-initio calculations based on many-body perturbation theory and the cumulant expansion with angle-resolved photoemission spectroscopy (ARPES) to quantify the electron-plasmon interaction in the highly-doped semiconducting transition metal dichalcogenide 1\textit{T}-HfS$_2$. ARPES reveals the emergence of satellite spectral features in the vicinity of quasiparticle excitations at the bottom of the conduction band, suggesting coupling to bosonic excitations with a characteristic energy of 200 meV. Our first-principles calculations of the photoemission spectral function reveal that these features can be ascribed to electronic coupling to carrier plasmons (doping-induced collective charge-density fluctuations). We further show that reduced screening at the surface enhances the electron-plasmon interaction and is primarily responsible for the emergence of plasmonic polarons. 
\end{abstract}
\maketitle
\section{Introduction}

The existence of satellite structures in the spectral function of solids has
been known since the infancy of photoemission spectroscopy \cite{baer_x-ray_1973}.  Satellites have first been identified by X-ray photoemission spectroscopy (XPS) in elemental metals -- such as Al \cite{baer_x-ray_1973,barrie_x-ray_1973}, alkali (Li, Na) \cite{kowalczyk_x-ray_1973}, and alkaline earth metals (Be, Mg)  \cite{pardee_analysis_1975} -- as broadened replica of the valence and core density of states  red-shifted by multiples of the plasmon energy.
Besides ordinary metals, photoemission satellites have been observed in pristine semiconductors (as, e.g., undoped silicon \cite{guzzo2011valence}) -- where the excitation of photoholes couples to valence plasmons.
Ab-initio calculations and angle-resolved photoemission experiments
later revealed that full band-structure replicas can arise from the
simultaneous excitations of a photohole and a valence plasmon \cite{caruso2015band,lischner_satellite_2015,caruso2015spectral}.

The interest in photoemission satellites has been revived by the discovery
of photoemission satellites due to the Fr\"ohlich electron-phonon
interactions in highly-doped anatase TiO$_2$ \cite{moser2013tunable}, 
and in the 2D electron gas formed at the surface of SrTiO$_3$ \cite{wang_tailoring_2016}. 
These features have been recognized as the smoking-gun evidence for the 
formation of Fr\"ohlich polarons -- strongly-coupled 
quasiparticles resulting from the dressing of photoexcited holes by polar longitudinal optical phonons \cite{franchini_polarons_2021,kang2018holstein}. 
Overall, the emergence of satellite structures in photoemission spectroscopy is a hallmark of strong electron-boson interaction in solids, and it has provided a strong stimulus for the development of new ab-initio theories for electron-boson coupling, including Fr\"ohlich coupling \cite{VerdiPRL2015,Verdi2017a,antonius2015dynamical}, density-functional \cite{pasquarello2018,sio_polarons_2019,gonzePRB2022} and many-body polaron theories \cite{lafuente-bartolome_unified_2022,franchini_polarons_2021}, electron-plasmon interaction \cite{caruso2016theory}, and the cumulant expansion approach \cite{Langreth1970,aryasetiawan1996multiple,guzzo2011valence,LischnerPRL2013,PhysRevB.90.085112,caruso2015band,lischner_satellite_2015,caruso2020many}. 

At variance with metals and pristine semiconductors,
satellites in highly-doped semiconductors and insulators occur
in the immediate vicinity of the band edges, and thus
influence fundamental properties of direct relevance for the
transport and dynamics of charge carriers, including quasiparticle
lifetimes and effective masses \cite{giustino2017electron,GonzePRB2021}. 
Changes of the doping concentration can further be exploited to exert control on the electron-phonon and electron-plasmon coupling strength, with visible effects on the structure of photoemission satellites
\cite{riley2018crossover}.
In particular, doping-induced free carriers can screen the electron-phonon interaction, suppressing the formation of Fr\"ohlich polarons and washing out the corresponding spectral fingerprints in ARPES \cite{moser2013tunable,Verdi2017a}. 
At the same time, at large doping concentrations   carrier plasmons can be excited in materials, with plasmon energies and electron-plasmon coupling strengths that increase with the carrier density. At strong coupling, electron-plasmon  interactions can result in the formation of plasmonic polarons with spectral signatures analogous to those of phonon-induced polaronic satellites \cite{riley2018crossover,ma_formation_2021,caruso2021two}.

Plasmonic polarons have thus far only been observed in a handful of materials, including EuO \cite{riley2018crossover}, anatase TiO$_2$ \cite{ma_formation_2021}, and monolayer MoS$_2$ \cite{caruso2021two}. 
A challenge that must be overcome for the observation of these phenomena consists in reaching the very high doping concentrations (of the order of $n=10^{20}~{\rm cm}^{-3}$) -- which are required for the emergence of an electron liquid while preserving the sample crystallinity.
In EuO,  these conditions have been realized via Eu-substitution by Gd \cite{riley2018crossover}; in anatase, TiO$_2$ free carriers are introduced by oxygen vacancies \cite{ma_formation_2021}; highly-doped MoS$_2$ monolayers have been realized by stimulating the formation of chalcogen vacancies via thermal annealing \cite{caruso2021two}. 

\begin{figure*}
    \centering
    \includegraphics[width=1.\linewidth]{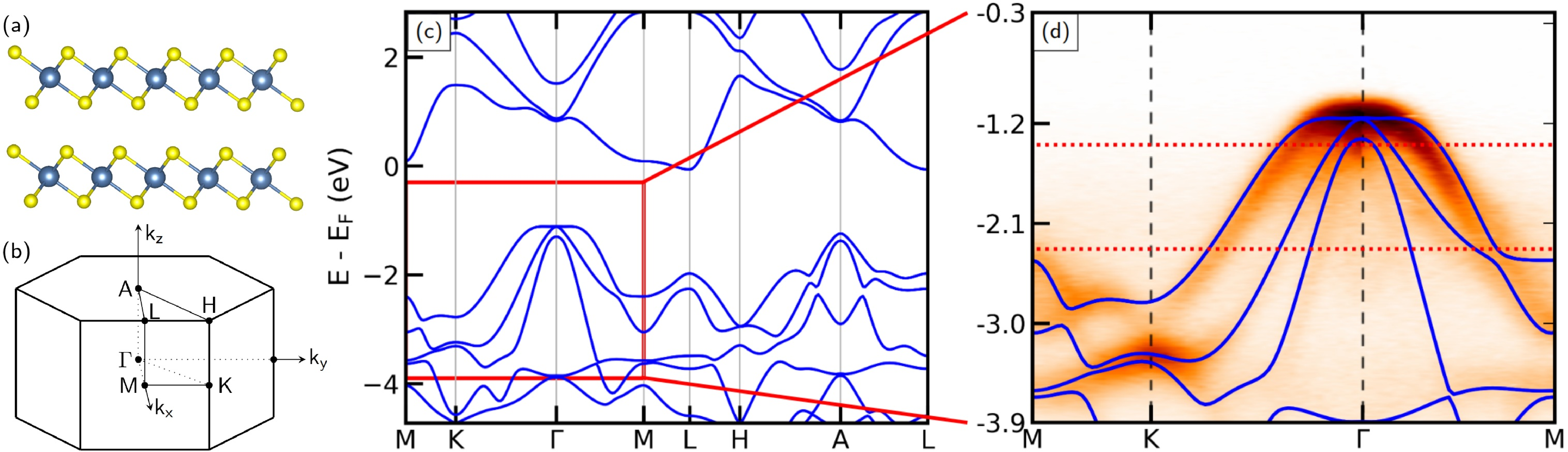}
    \caption{(a) Crystal structure of 1\textit{T}-HfS$_2$ (side view) and (b) corresponding hexagonal Brillouin zone. (c) DFT band structure of pristine bulk 1\textit{T}-HfS$_2$. Energies are relative to the conduction band minimum. (d) DFT band structure superimposed on the ARPES spectral function for energies and crystal momenta within the red rectangle in panel (c). Dark (light) colors denote high (low) values of the spectral function. The dashed red lines mark the energies of the ARPES intensity maps of Fig.~\ref{fig:2}(a)-\ref{fig:2}(b).}
    \label{fig:1}
\end{figure*}

In this work, we realize strong electron-plasmon interactions via the deposition of alkali atoms on the surface of hafnium disulfide (1\textit{T}-HfS$_2$). 
To corroborate this new way of controlling the electron-plasmon interaction, we conduct a combined theoretical and experimental investigation of the electronic and quasiparticle excitations for pristine and highly-doped 1\textit{T}-HfS$_2$. ARPES measurements for $n$-doped samples reveal the emergence of satellite spectral structures in the vicinity of the quasiparticle peak at the bottom of the conduction band. 
To unravel the origin of these features we performed ab-initio calculations of the electron spectral function by explicitly including the influence of electron-plasmon interaction in the Fan-Migdal approximation. Spectral function calculations based on the cumulant expansion approach -- the state of the art for the description of satellites in photoemission -- are in excellent agreement with ARPES experiments, corroborating the plasmonic origin of the ARPES satellites. These findings demonstrate that alkali doping in bulk transition metal dichalcogenides can alter the spectrum of quasiparticle excitations, providing a viable route to realize strong electron-plasmon coupling.

The manuscript is structured as follows. In Sec.~\ref{sec:methods} experimental and computational methods are discussed. In Sec.~\ref{sec:pristineHfS2}, we present ARPES measurements and ab-initio calculations of pristine 1\textit{T}-HfS$_2$. In Sec.~\ref{sec:Polarons} we discuss the theory and measurements of plasmonic polarons in the ARPES spectral function of highly-doped 1\textit{T}-HfS$_2$. Concluding remarks are presented in Sec.~\ref{sec:conc}. 

\section{METHOD}
\label{sec:methods}

1\textit{T}-HfS$_2$ single crystals were grown by chemical vapor transport at the in-house facilities. The sample was cleaved inside the ultra-high vacuum chamber at room temperature and subsequently transferred to the liquid helium-cooled manipulator for photoemission measurements. During the ARPES measurements, the sample temperature was maintained at 10 K. {\it In situ} doping of the 1\textit{T}-HfS$_2$ samples was achieved by depositing potassium atoms from an alkali metal dispenser (SAES Getters) on the surface. The dopant atoms adsorbed on the surface and sub-surface, but did not intercalate into deeper layers of the van der Waals material.

The experiments were performed at beamline P04 of PETRA III at DESY using the ASPHERE photoelectron spectroscopy endstation. The area probed by the synchrotron beam had a size of approximately $15\times15~\mu{\rm m}^2$, the photon energies used and corresponding total energy resolution of the ARPES measurements were within a range of 260-450~{eV} and 50-80~meV, respectively. The Fermi surface map of the doped 1\textit{T}-HfS$_2$ sample in Fig.~\ref{fig:2} was recorded at a photon energy of 432~eV, probing the 11th $\Gamma$ point in the $k_z$ direction.

Density functional theory (DFT) calculations were performed with the plane-wave pseudopotential code {\tt Quantum ESPRESSO} \cite{giannozzi2009quantum}. We used the Perdew-Burke-Ernzerhof (PBE) generalized gradient approximation for the exchange-correlation functional \cite{PhysRevLett.77.3865} and Optimized Norm-Conserving Vanderbilt fully relativistic pseudopotentials  \cite{PhysRevB.88.085117}. The plane-wave kinetic energy cutoff was set to 120~Ry and the integrals over the Brillouin zone were discretized on a $12\times 12 \times 6$ Monkhorst-Pack $k$-point mesh. Spin-orbit coupling (SOC) was included at all steps of our calculations. 
The band structure was interpolated onto a $\mathrm{60 \times 60 \times 30}$ homogeneous grid via maximally-localized Wannier functions \cite{MLWF12} as implemented in the {\tt WANNIER90} package \cite{pizzi2020wannier90}. 
The effect of $n$-type doping was included by rigidly shifting the Fermi level above the conduction-band bottom to account for additional free carriers. Charge neutrality of the system is retained by introducing a compensating positively charged homogeneous background.
Ab-initio calculations of the electron-plasmon interaction were conducted with the {\tt EPW} code \cite{ponce2016epw} and employed the Fan-Migdal approximation for the electron self-energy and the cumulant expansion for the spectral function \cite{guzzo2011valence,caruso2016gw}. 

\section{Electronic properties of pristine 1\textit{T}-HfS$_2$}
\label{sec:pristineHfS2}

\begin{figure*}
    \centering
    \includegraphics[width=1\textwidth]{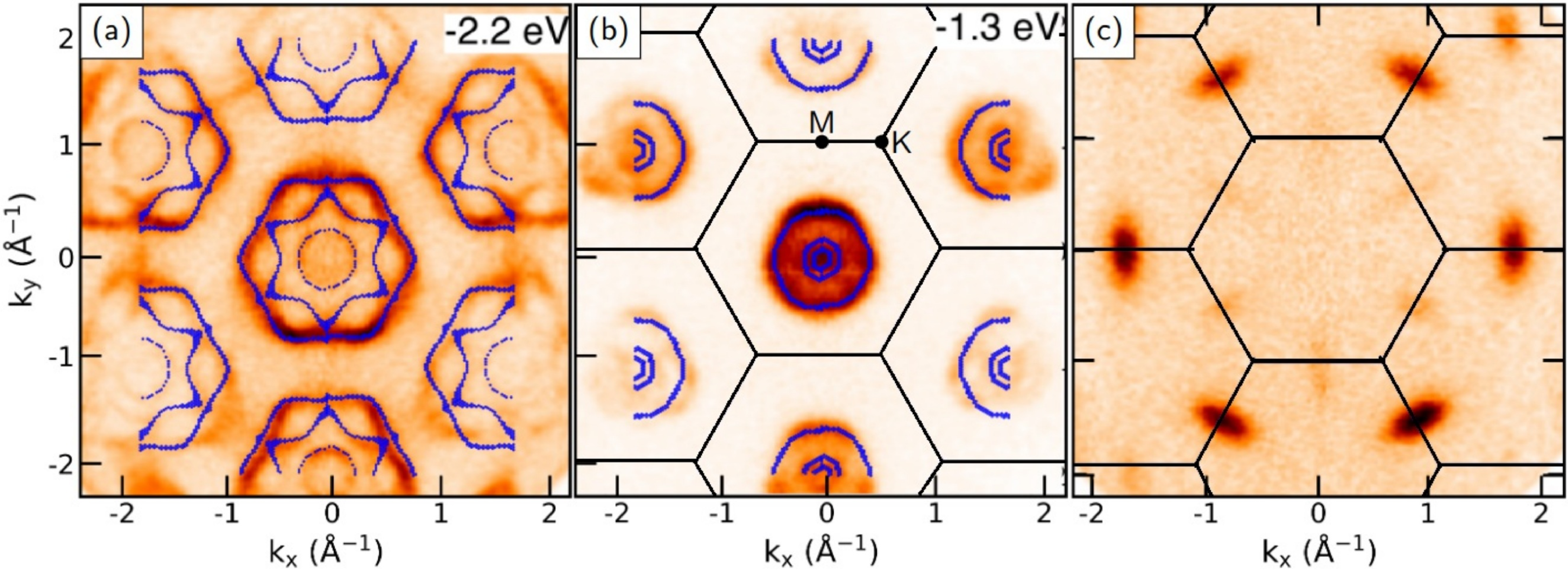}
    \caption{ARPES intensity distributions in the ${k_x}$-${k_y}$-plane taken at -2.2\,eV (a) and -1.3\,eV (b) relative to the conduction band minimum for pristine 1\textit{T}-HfS$_2$. The DFT bands are marked in blue. Black hexagons mark the boundaries of the Brillouin zone with the $\mathrm{\Gamma}$ point in the centre. The position of the M and K high-symmetry points is marked in panel (b). (c) ARPES intensity map taken at the Fermi level of highly-doped 1\textit{T}-HfS$_2$. Ellipsoidal intensity patterns reflect the population of the bottom of the conduction band by alkali deposition-induced free-carrier doping. Due to ARPES matrix element effects the intensity of the pockets in the first Brillouin zone is suppressed.}
    \label{fig:2}
\end{figure*}

1\textit{T}-HfS$_2$ crystallizes in a layered crystal structure with a hexagonal unit cell belonging to the 164 space group ($P\overline{3}m1$). A side view of the 1\textit{T}-HfS$_2$ crystal structure and its Brillouin zone are shown in Fig.~\ref{fig:1}(a) and \ref{fig:1}(b), respectively. The bulk band structure of pristine 1\textit{T}-HfS$_2$ as obtained from DFT-PBE is shown in Fig.~\ref{fig:1}(c). The path across the Brillouin zone passes through the M-K-$\Gamma$-M and the L-H-A-L high-symmetry points and it was chosen to facilitate comparison with the ARPES measurements. 1\textit{T}-HfS$_2$ is an indirect band gap semiconductor with the valence band maximum (conduction band minimum) located at the $\Gamma$ (L) high-symmetry point. The calculated indirect band gap of 1.2~eV is in good agreement with earlier DFT studies \cite{iordanidou2016impact,shang2017electric}. 
Analysis of the projected DOS (not shown) reveals that the valence bands arise primarily from the hybridization of $p$-orbitals with S character, while the conduction bands are predominantly characterized by the $d$-orbitals with Hf character \cite{neal2021}. In heavy elements with unfilled 5d orbitals, such as Hf, SOC has important effects on the electronic structure \cite{lau2019electronic}. In 1\textit{T}-HfS$_2$ it leads to a shift of the valence band maximum to $\Gamma$ and induces a bandsplitting of the two highest valence bands, while the conduction band minimum remains unaffected. The influence of SOC on the band structure is further discussed in Appendix~A.

A parabolic fitting to the conduction-band minimum along the three reciprocal lattice vectors yields the following values for the electron effective masses $m^*_1= 0.25 ~m_e, m^*_2= 1.65 ~m_e, \text{ and } m^*_3= 0.20 ~m_e$, which are in good agreement with earlier DFT calculations \cite{lu2018band}. The density of states (DOS) effective mass was determined as $m^*_{\mathrm{DOS}} = (g m^*_{1}m^*_{2}m^*_{3} )^{3/2}$, where $g=6$ is the degeneracy factor of the conduction band minimum, yielding $m^*_{\mathrm{DOS}} = 1.44 \,m_{e}$ \cite{green1990intrinsic}.
 
The measured ARPES spectral function for the valence band along the M-K-$\Gamma$-M high-symmetry path is shown in Fig.~\ref{fig:1}(d).  The DFT-PBE band structure, superimposed on the measurements for comparison, is in very good agreement with the experiments. We observe a small deviation between measurements and calculations in the vicinity of the K high-symmetry point for energies around -3.0~eV,  which we tentatively attribute to the finite ${k_z}$ broadening, resulting in the superposition of ARPES intensities corresponding to different ${k_z}$-planes of the Brillouin zone.

In Figs.~\ref{fig:2}(a) and \ref{fig:2}(b) we show the measured ARPES intensity maps for crystal momenta spanning the ${k_x}$-${k_y}$-plane for energies corresponding to -2.2\,eV and -1.3\,eV below the Fermi energy marked by horizontal dashed lines in Fig.~\ref{fig:1}(d), respectively. The ab-initio band structure evaluated from DFT and interpolated using maximally-localized Wannier functions closely matches the experimental data within the first Brillouin zone. The slight deviations in the second Brillouin zones are  attributed to the $k_z$ variation of constant-energy ARPES angle maps.

\section{Polarons in highly-doped 1\textit{T}-HfS$_2$}
\label{sec:Polarons}

\begin{figure*}
    \centering
    \includegraphics[width=1\textwidth]{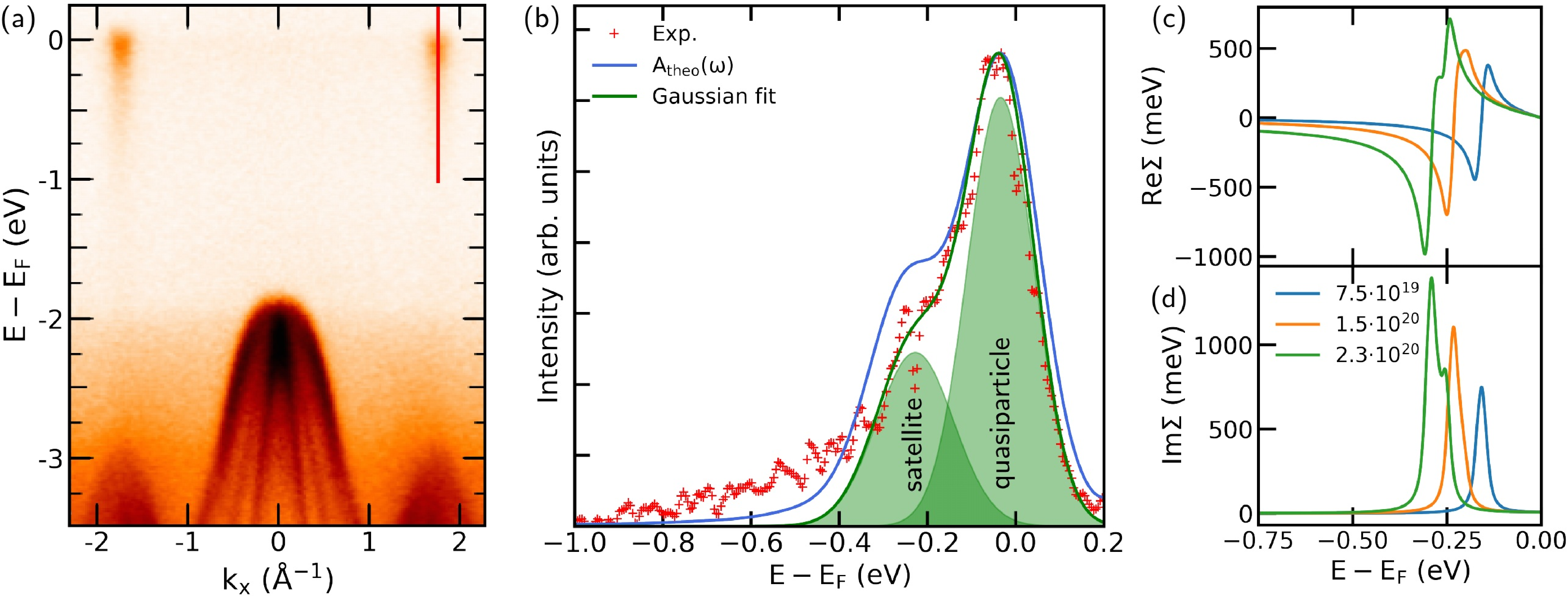}
    \caption{(a) ARPES measurements of highly-doped 1\textit{T}-HfS$_2$ with a charge carrier concentration of $n = 1.5\cdot 10^{20} \mathrm{cm^{-3}}$. ARPES intensity map around the Fermi level reflecting the population of the conduction-band bottom and exhibiting a polaronic tail extending down to -1\,eV below the conduction band edge. (b) ARPES spectrum (red crosses) and {\it ab-initio} spectral function (blue line) in the vicinity of the conduction-band bottom for energies and momentum marked by the red line in panel (a). A Shirley background was subtracted from the experimental data. The spectral function calculations are based on cumulant expansions and broadened by a Gaussian with a variance of 80\,meV to match the experimental resolution. Two Gaussians (green shading) illustrate the decomposition of the total ARPES intensity into a quasiparticle peak and a satellite peak. (c)-(d) Real and imaginary parts of the electron self-energies due to electron-plasmon interaction for different carrier concentrations given in cm$^{-3}$.} 
    \label{fig:3}
\end{figure*}

In the following we investigate the influence of $n$-type doping on the 
band structure and on the spectrum of quasiparticle excitations of 1\textit{T}-HfS$_2$. 
Figure~\ref{fig:2}(c) shows the ARPES measurement of the Fermi surface of highly-doped 1\textit{T}-HfS$_2$. While no signal is seen at these energies for the pristine sample, finite intensity arises from the population of the conduction band due to doping. The elliptical intensity pattern reflects the anisotropic band dispersion of the lowest conduction band. Due to photoemission matrix element effects, the intensity in the first Brillouin zone is suppressed. The extrinsic carrier concentration of $n =1.5\cdot 10^{20}\,$~cm$^{-3}$ is extracted from the size of the Fermi pockets of Fig.~\ref{fig:2}(c). 

The extrinsic charge carriers introduced by $n$-type dopants can significantly modify the electronic properties \cite{ziambaras2007potassium,rossnagel2010suppression}. In addition to the population of the conduction-band bottom, the doping-induced extrinsic carriers can lead to the emergence of carrier plasmons with a characteristic frequency given by ${\omega\drm{pl} = \sqrt{4 \pi {n} e^2/{ m_{\rm DOS}^{*}} \epsilon\drm{\infty}}}$, with high-frequency dielectric constant $\epsilon\drm{\infty}$ and DOS effective mass  $m_{\rm DOS}^{*}$ \cite{giuliani2005quantum}. In doped semiconductors, the plasmon energy can span values between 10 and 200~meV \cite{caruso2016theory}. These low-energy plasmons can further couple to carriers in the conduction band via electron-plasmon interactions leading to the emergence of polaronic quasiparticle excitations \cite{riley2018crossover,caruso2018electron,ma_formation_2021,caruso2021two}. In the following, we proceed to investigate these phenomena on a quantitative ground, by combining ab-initio theory and ARPES measurements of highly-doped 1\textit{T}-HfS$_2$.

The band structure of the $n$-doped 1\textit{T}-HfS$_2$ obtained by ARPES along $\Gamma$-K-M direction is shown in Fig.~\ref{fig:3}(a). The energies are relative to the Fermi level, which is located 50~meV above the conduction-band bottom. Compared to the pristine sample, structural disorder and additional doping-induced scattering processes contribute to an enhancement of the band structure broadening. For photoelectron energies above the fundamental gap (1.9~eV), our measurements reveal the emergence of additional photoemission intensities that  reflect the population of the conduction-band bottom by the alkali deposition-induced carriers. The Fermi pockets of the conduction band are centered around ${|k_{x}| = 1.8\,}${\AA}$^{-1}$.
Figure~\ref{fig:3}(b) illustrates the ARPES spectral function $A\drm{exp}(\omega)$ for energies and crystal momentum marked by the red line in Fig.~\ref{fig:3}(a). In Fig.~\ref{fig:3}(b),  we eliminated the background signal from the experimental data by subtracting a Shirley background function $B(\omega)$, defined as  ${ B(\omega) = \beta \int^\mu_\omega d\omega' I(\omega') }$  where $\beta$ is an adjustable parameter and $\mu$ is the chemical potential \cite{shirley1972high}. The resulting spectral function $A\drm{exp}(\omega)$ is characterized by a sharp quasiparticle peak and an additional shoulder structure  at 200~meV below the Fermi level, with a decreasing photoemission intensity extending down to 1~eV below the Fermi level. A Gaussian decomposition of the ARPES intensity, marked in green in Fig.~\ref{fig:3}(b), suggests that these spectral features are compatible with a superposition of a quasiparticle peak and a photoemission satellite peak red-shifted by 200~meV from the maximum of the quasiparticle peak. 

The emergence of photoemission satellites in doped semiconductors is a 
hallmark of strong electron-boson interaction which has been widely 
investigated in the past owing to its close relation to the formation 
of Fr\"ohlich polarons --  a prototypical emergent phenomenon due to strong electron-phonon coupling. The energy separation between quasiparticle and satellite peaks 
is expected to match the energy of the boson that underpins the coupling. 
For example, Fr\"ohlich polarons in polar semiconductors arise 
from the coupling of $n$-type carriers with 
polar longitudinal-optical  (LO) phonons,  and they manifest themselves in ARPES spectra 
via satellite structures at energies matching the LO phonon energies. 
In 1\textit{T}-HfS$_2$, the energy separation between the quasiparticle and satellite peak (200~meV) exceeds  
the LO phonon energies ($<$40~meV,  see e.g., the phonon dispersion in Appendix B). 
The discrepancy of these energy scales enables us to promptly exclude the Fr\"ohlich 
electron-phonon interaction as a source of polaronic coupling. 
The absence of spectral fingerprints of Fr\"ohlich polarons
can be easily rationalized by noting that (i) 1\textit{T}-HfS$_2$ is a weakly 
polar crystal, i.e., it is characterized by small Born effective 
charges, and (ii) at the high doping concentration considered in 
our work electron-phonon coupling is screened by free carriers, thus, 
further mitigating the effects of Fr\"ohlich coupling. 

In the following, we thus proceed to inspect the  electron-plasmon interaction as 
a possible source of polaronic coupling, and we analyze its influence on the emergence of 
photoemission satellites.
To quantify the electron-plasmon interaction and its effect on the ARPES measurements, we 
evaluate the electron self-energy due to the electron-plasmon interaction, which in the Fan-Migdal approximation can be expressed as \cite{caruso2016theory}:
\begin{align}
     & \Sigma_{n \mathbf{k}}\urm{epl} = \int \frac{d \mathbf{q}}{\Omega\drm{\rm BZ}} \sum_{m} |g_{mn}\urm{epl}(\mathbf{k},\mathbf{q})|^2  \label{eq:sigmaepl}\\ 
     \times&  \left[ \frac{n\drm{\mathbf{q}}+f_{m \mathbf{k} + \mathbf{q}}}{\varepsilon_{n \mathbf{k}} - \varepsilon_{m \mathbf{k} + \mathbf{q}} + \hbar \omega\urm{pl}\drm{\mathbf{q}} + i \eta}  
     + \frac{n\drm{\mathbf{q}} +1 - f_{m \mathbf{k} + \mathbf{q}}}{\varepsilon_{n \mathbf{k}} - \varepsilon_{m \mathbf{k} + \mathbf{q}} -  \hbar \omega\urm{pl}\drm{\mathbf{q}} + i \eta} \right]
     \nonumber
\end{align}
where $\Omega\drm{BZ}$ is the Brillouin zone volume, $m$ and $n$ are band indices, $\mathbf{k}$ and $\mathbf{q}$ are Bloch wave vectors, $n\drm{\mathbf{q}}$ denotes the Bose-Einstein and $f_{m \mathbf{k} + \mathbf{q}}$ the Fermi-Dirac distribution, $\varepsilon$ are the Kohn-Sham (KS) eigenstates, $\omega\urm{pl}\drm{\mathbf{q}}$ is the plasmon frequency and $\eta$ is a positive infinitesimal. The integral runs over the Brillouin zone volume. The first term accounts for electron scattering processes involving the absorption of a plasmon $+\omega\urm{pl}\drm{\mathbf{q}}$, while the second term accounts for hole scattering processes mediated by plasmon emission. $g_{mn}\urm{epl}$ denotes the electron-plasmon coupling matrix elements, that can be expressed as \cite{caruso2016theory}:
\begin{align}
    g_{mn}\urm{epl}(\mathbf{k},\mathbf{q}) &= \left[ \left.\frac{\partial\epsilon(\mathbf{q},\omega)}{\partial \omega} \right|_{\omega\urm{pl}_{\mathbf{q}}} \right]\urm{-\frac{1}{2}} \label{eq:gammaepl}\\
    &\times\left(\frac{4\pi}{\Omega\drm{BZ}}\right)^{\frac{1}{2}} \frac{1}{|\mathbf{q}|} 
    \bra{\psi_{m \mathbf{k} + \mathbf{q}}} e{ i \mathbf{q}\cdot \mathbf{r}}\ket{\psi_{n \mathbf{k}}}. \nonumber 
\end{align}
Here, $\epsilon$ is the dielectric function, $\bra{\psi_{m \mathbf{k} + \mathbf{q}}} e{ i \mathbf{q}\cdot \mathbf{r}}\ket{\psi_{n \mathbf{k}}}$ the dipole matrix element and $\psi_{\mathbf{k}}$ the Kohn-Sham orbital. The  $|\mathbf{q}|^{-1}$ singularity in the electron-plasmon coupling matrix element is reminiscent of the Fr\"ohlich interaction in polar semiconductors, and it indicates that the long-wavelength plasmons dominate electron-plasmon scattering processes.

Owing to the dependence of the matrix elements $g_{mn}\urm{epl}$ on the 
dielectric function, the electron-plasmon interaction is profoundly 
influenced by the screening environment of the system. In 1\textit{T}-HfS$_2$, the alkali 
dopant atoms are concentrated in the vicinity of the surface and, possibly, underneath the first 
1\textit{T}-HfS$_2$ layers of the sample.
Correspondingly, the dielectric screening experienced by $n$-type carriers
is mitigated as compared to bulk carriers. 
To account for the charge localization at the surface we introduce an effective dielectric constant $\mathrm{\varepsilon\urm{S}\drm{\infty} = (\varepsilon\urm{HfS_2}\drm{\infty} +1)/2} = 3.6$, with $\mathrm{\varepsilon\urm{HfS_2}\drm{\infty} = 6.2}$ being the high-frequency dielectric constant of bulk 1\textit{T}-HfS$_2$ \cite{lucovsky1973infrared, iwasaki1982anisotropy}. Further details on the evaluation of the electron-plasmon matrix elements can be found elsewhere \cite{caruso2018electron}.
Based on this value, we estimate the plasmon frequency to be 200~meV for a doping concentration $n=1.5\cdot10^{20}~{\rm cm}^{-3}$, which matches closely the satellite energy, thus, suggesting electron-plasmon coupling as a likely origin of this polaronic feature.

In Figs.~\ref{fig:3}(c)-(d) the real and imaginary part of the electron self-energy due to electron-plasmon coupling are presented, respectively, for doping carrier concentrations $n=7.5\cdot 10^{19}$, $1.5\cdot10^{20}$, and $2.25\cdot 10^{20}$~cm$^{-3}$. The middle value coincides with the doping concentration determined from experiment. 
The corresponding imaginary parts of the self-energy in Fig.~\ref{fig:3}(d) exhibit a sharply peaked structure with a Lorentzian line profile in the vicinity of the energy ${\varepsilon_{\mathbf{k}} - \hbar \omega\drm{pl}}$. For larger doping concentrations, we observe a progressive red-shift  of the peak in ${\rm Im} \Sigma$ and an increase of its intensity, which arise from the increase of plasmon energy and of the electron-plasmon coupling matrix elements, respectively.
The real part of the self-energy is related to ${\rm Im} \Sigma$ by a 
Kramers-Kronig's transformation and it thus also has a similar dependence on the doping concentration. 

Based on the electron self-energy, we proceed to investigate the signatures of electron-plasmon coupling in ARPES.
Earlier studies revealed that ab-initio calculations of photoemission satellites based on the Fan-Migdal approximation overestimate the satellite energy and intensity by up to 50\%  as compared to experiment \cite{guzzo2011valence,caruso2020many}. To circumvent this limitation, we evaluate the spectral function  based on the cumulant expansion approach \cite{guzzo2011valence}. The cumulant expansion representation of the spectral function  can be expressed as \cite{aryasetiawan1996multiple,aryasetiawan1998gw,caruso2020many}:
\begin{equation}
    A(\mathbf{k},\omega) = \sum_n e^{A\urm{S1}_{n\mathbf{k}}(\omega)} * A\urm{QP}_{n\mathbf{k}}(\omega).
    \label{eq:A(w)}
\end{equation}
Here, $*$ denotes a convolution product and $A\urm{QP}_{n\mathbf{k}}(\omega) = 2\pi^{-1} \operatorname{Im}[\hbar \omega - \varepsilon_{n \mathbf{k}} - \Sigma_{n \mathbf{k}}\urm{epl}(\varepsilon_{n \mathbf{k}})]^{-1}$ is the quasiparticle contribution to the spectral function evaluated in the \enquote{on the energy shell} approximation, in which the frequency dependence of the self-energy $\Sigma_{n \mathbf{k}}(\omega)$ is replaced by the KS energy $\omega = \varepsilon_{n\mathbf{k}}$ \cite{gumhalter2016combined, quinn1958electron}. 
The satellite part of spectral function is further given by \cite{caruso2018electron}:
\begin{align}
    A\urm{S1}_{n\mathbf{k}}(\omega) = - \frac{\beta_{n\mathbf{k}}(\omega) - \beta_{n\mathbf{k}}(\varepsilon_{n\mathbf{k}}) - (\omega - \varepsilon_{n\mathbf{k}}) \beta'_{n\mathbf{k}}(\varepsilon_{n\mathbf{k}})}{(\omega - \varepsilon_{n\mathbf{k}})^2},     \label{eq:A_S(w)}
\end{align}
with $\beta = \pi^{-1}\operatorname{Im} \Sigma_{n \mathbf{k}}(\varepsilon_{n\mathbf{k}} - \omega) \Theta(\omega)$ and $\beta'$ denoting its first derivative. The first term in the Taylor series expansion of the exponential in Eq.~\eqref{eq:A(w)} corresponds to the quasiparticle peak of the photoemission spectrum, while higher-order terms  account for plasmon-assisted scattering up to infinite order. In the following, we truncated Eq.~\eqref{eq:A(w)} to the second order. Terms above the second order contribute negligibly to the spectral function and their inclusion is inconsequential. 

The spectral function in the vicinity of the conduction-band bottom computed from Eqs.~\eqref{eq:A(w)}-\eqref{eq:A_S(w)} is shown in Fig.~\ref{fig:3}(b). To account for the finite experimental resolution, the spectral function has been convoluted with a Gaussian with a variance of 80~meV. The intensity of the convoluted spectral function is rescaled to match the experimental spectrum at the quasiparticle peak. Experimental broadening and intensity rescaling are the only adjustable parameters in our simulations.
In short, the cumulant spectral function exhibits a pronounced shoulder arising from the convolution of the satellite and quasi-particle spectral function $A\urm{S} * A\urm{QP}$. This spectral feature corresponds to the coupled excitation of a photohole and a plasmon and matches closely the photoemission satellite measured in ARPES. Higher-order satellite structures due to multiple plasmon excitations have small intensity and they are washed out by the finite experimental resolution. 
Overall, the close agreement between simulations and 
measurements suggest that carrier plasmons in highly-doped 1\textit{T}-HfS$_2$ 
are strongly coupled to free carriers in the conduction bands, 
leading to the formation of plasmonic polarons and to corresponding 
spectral fingerprints in the ARPES spectrum. 
The residual discrepancy between theory and experiments at energies 
smaller than -0.5~eV is tentatively attributed to impurity scattering and statistical noise, 
which are not captured by the Shirley background model.

\section{Conclusion}\label{sec:conc}

In summary, we conducted ab-initio calculations and ARPES measurements of the electronic properties and quasiparticle excitation of pristine and highly-doped 1\textit{T}-HfS$_2$.
We report the observation of polaronic satellites in the ARPES spectral function, which we attribute to the formation of plasmonic polarons. Our first-principles calculations of the Fan-Migdal self-energy for electron-plasmon interaction explicitly account for extrinsic carriers introduced by alkali doping and closely reproduce the spectral fingerprints of polaronic satellites in the measured ARPES spectral function. 
In particular, the alkali doping enables the injection of free carriers in the vicinity of the surface, where screening is weak and it thus provides ideal conditions for realizing strong coupling between free carriers and plasmons.

Overall, our combined theoretical and experimental investigation reveals the possibility to tailor quasiparticle excitations and the electron-plasmon coupling strength via extrinsic doping mediated by the adsorption and  intercalation of alkali atoms affecting the first atomic layers of 1\textit{T}-HfS$_2$.

Materials interfaces and hybrid heterostructures may provide further opportunities to directly control the dielectric environment and, thus, tailor the spectrum of quasiparticle interactions.

\section*{ACKNOWLEDGMENTS}
This work was funded by the Deutsche Forschungsgemeinschaft (DFG), Projects No. 499426961 and 434434223 -- SFB 1461.
We thank DESY (Hamburg, Germany), a member of the Helmholtz Association HGF, for the provision of experimental facilities. Parts of this research were performed at PETRA III. Funding for the photoemission spectroscopy instrument at beamline P04 (contracts 05KS7FK2, 05K10FK1, 05K12FK1, and 05K13FK1 with Kiel University; 05KS7WW1 and 05K10WW2 with the University of Würzburg) by the German Federal Ministry of Education and Research (BMBF) is gratefully acknowledged.

\section*{Appendix A: Influence of spin-orbit coupling on the band structure}
\label{sec:SOC}
\begin{figure}[h!]
    \centering
    \includegraphics[width=1\linewidth]{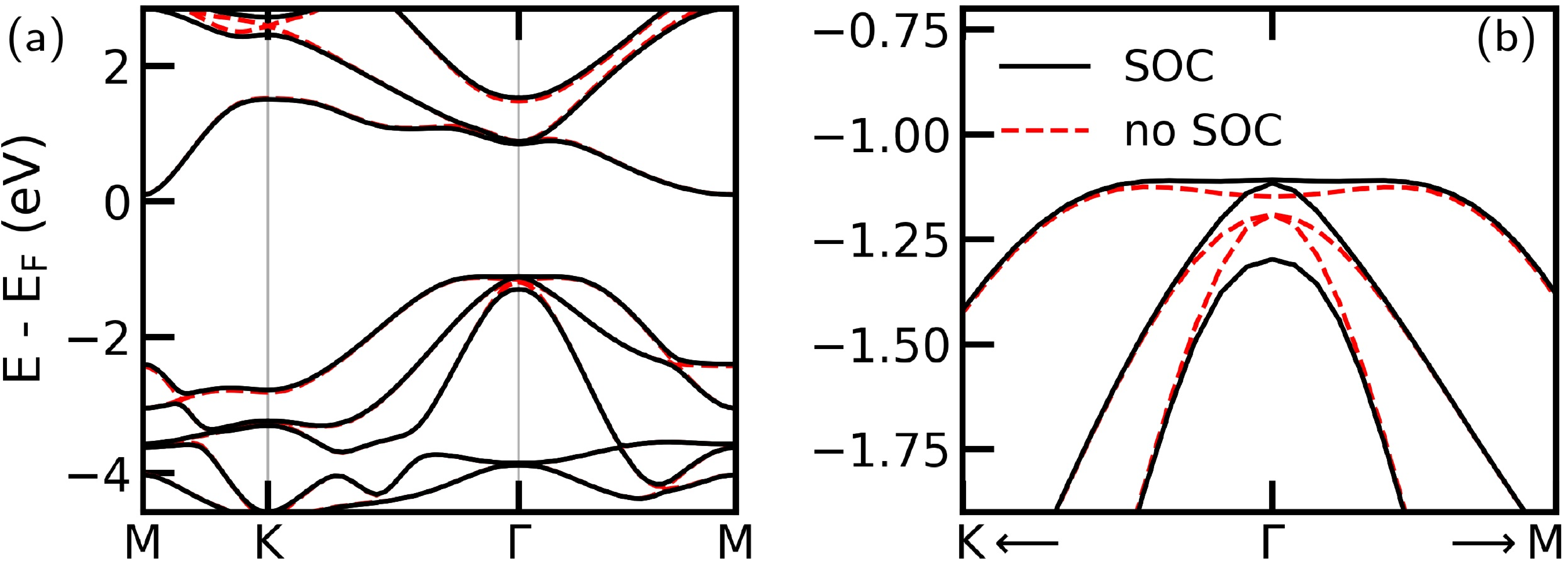}
    \caption{Electronic band structure of 1\textit{T}-HfS$_2$ calculated using DFT with and without SOC.}
    \label{fig:SOC}
\end{figure}
In Fig.~\ref{fig:SOC} the band structure of 1\textit{T}-HfS$_2$ with (black, continuous) and without (red, dashed) SOC is depicted. SOC leads to an avoided crossing in both the valence and conduction bands, lifting the degeneracy of upper valence bands at the $\mathrm{\Gamma}$ point. These findings are in good agreement with earlier calculations \cite{lau2019electronic}. 

\section*{Appendix B: Phonon dispersion}
\label{sec:PHDISP}
\begin{figure}[h!]
    \centering
    \includegraphics[width=1\linewidth]{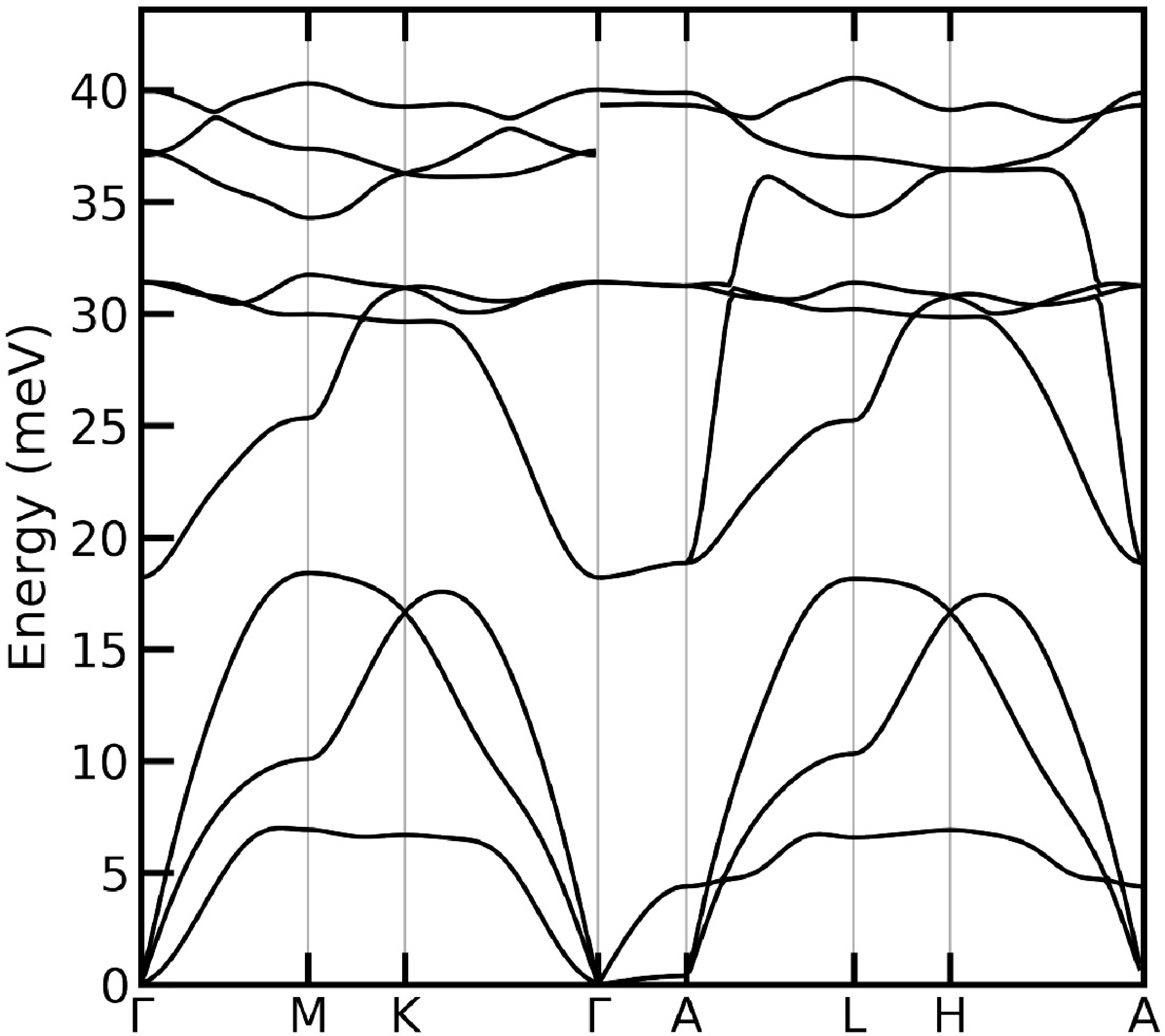}
    \caption{Phonon dispersion of 1\textit{T}-HfS$_2$ calculated using DFPT.}
    \label{fig:phonondisp}
\end{figure}
In Fig.~\ref{fig:phonondisp} the phonon dispersion of 1\textit{T}-HfS$_2$ obtained from density functional perturbation theory (DFPT) is presented. The discontinuity of the second and third highest-energy modes at the $\Gamma$ point arises from the LO-TO splitting. The highest phonon mode has a frequency of 43~meV. 
%
%\bibliography{references}

\end{document}